\documentstyle[epsf,12pt]{article}
\twocolumn
\newcommand{\be}{\begin{equation}}
\newcommand{\ee}{\end{equation}}
\newcommand{\ba}{\begin{eqnarray}}
\newcommand{\ea}{\end{eqnarray}}

\begin{document}
\begin{titlepage}
\pagestyle{empty}
\rightline{BUHEP-94-34}
\rightline{Dec. 1994}
\vskip .4in
\begin{center}
{\Large\bf   Numerical studies of instanton induced baryon
decay\footnote{Contribution to the Proceeding of the 1994 International
Symposium on Lattice Gauge Theories, held in Bielefeld, Germany,
Sep. 27 -- Oct. 1, 1994.  } }
\end{center}

\vskip .1in
\begin{center}
Claudio Rebbi\footnote{email: rebbi@pthind.bu.edu}

and

Robert Singleton, Jr.\footnote{email: bobs@cthulu.bu.edu }
\vskip.2in
{\it Department of
Physics, Boston University, \\ Boston, MA 02215, USA}

\vskip .2in
\end{center}

\vskip 1in
\centerline{ {\bf Abstract} }
\baselineskip=18pt
We describe an application of computational techniques to the
study of instanton induced baryon decay in high energy electroweak
interactions.
\noindent
\end{titlepage}
\baselineskip=18pt

We wish to illustrate an application of numerical methods to
the semiclassical study of instanton induced baryon number violation.
In this investigation, computational tools take over where more
traditional analytic methods reach their limit, and the two approaches
complement one another.  Our presentation will be in the form of a
progress report since the computation offers formidable challenges
and we have not yet brought it to completion. We believe, however,
that we have made enough progress to warrant the presentation of our
interim results, both because of the intrinsic interest of the technique
and in the hope that it may serve as inspiration for similar computational
investigations.

Since the pioneering work of 't Hooft \cite{thooft76} it has been
known that electroweak interactions can give origin to baryon number
violation because of the anomalous divergence of the baryon
current and the occurrence of topology changing instanton processes.
Nevertheless, at low energy and low temperature the amplitude for
baryon non-conserving processes is abysmally small because of the
barrier penetration factor:

\be
A \approx \exp(-S_{inst} / g^2) \ , \label{eq1}
\ee

\noindent
where $g$ is the electroweak coupling constant and $S_{inst}$, the
action of the instanton configuration in rescaled units, is a
numerical factor of order one.  It has been shown that the barrier
can be overcome by thermal fluctuations if the temperature becomes
comparable to the energy $E_{sph}$ of the sphaleron, the metastable
state which sits on top of the barrier separating field configurations
of different topology.  A much more controversial question is whether
unsuppressed baryon number violation can also occur in high energy
collisions when $E \approx E_{sph} $. The difficulty stems from the
fact that one needs to consider processes where the initial state is
an exclusive two particle state. So far semiclassical techniques have
provided the only clues to the study of instanton induced baryon number
violation, but these methods are not amenable to the study of exclusive
state processes. Recently a framework for solving the problem has been
proposed by Rubakov, Son and Tinyakov \cite{rst92}. (The literature on
instanton induced baryon number violation is very rich, and although
many authors deserve credit, limitations of space force us to cite
only work directly related to our investigation.) The main idea
is to consider processes
where the initial state is also inclusive, but constrained so that
the particle number operator takes a fixed value $N=\nu / g^2$.
The total energy is given by $E=\epsilon / g^2$ and, with $\nu$
and $\epsilon$ held fixed, the small $g$ limit allows one to
evaluate functional integrals by a saddle point approximation.
This leads to a cross section \mbox{$\sigma \approx \exp[-F(\epsilon, \nu) /
g^2]$}, and the the limiting value of $F$ for $\nu \to 0$ provides
information for processes with small numbers of initial particles.

With some non-trivial manipulations one can determine the saddle point
configuration that dominates the functional integral \cite{rst92}.
This is given by the solution of the complexified classical equations
of motion along a special contour in the complex time plane, shown
in Fig. 1. In general such a solution cannot be found analytically,
but its numerical calculation is within reach of today's computers.
This is especially true if one considers either a simplified model,
or configurations with rotational symmetry reducing the degrees of
freedom to one-space and one-time dimensions.
\vspace{3mm}

\epsfxsize=75mm
\epsfysize=85mm
\epsfbox{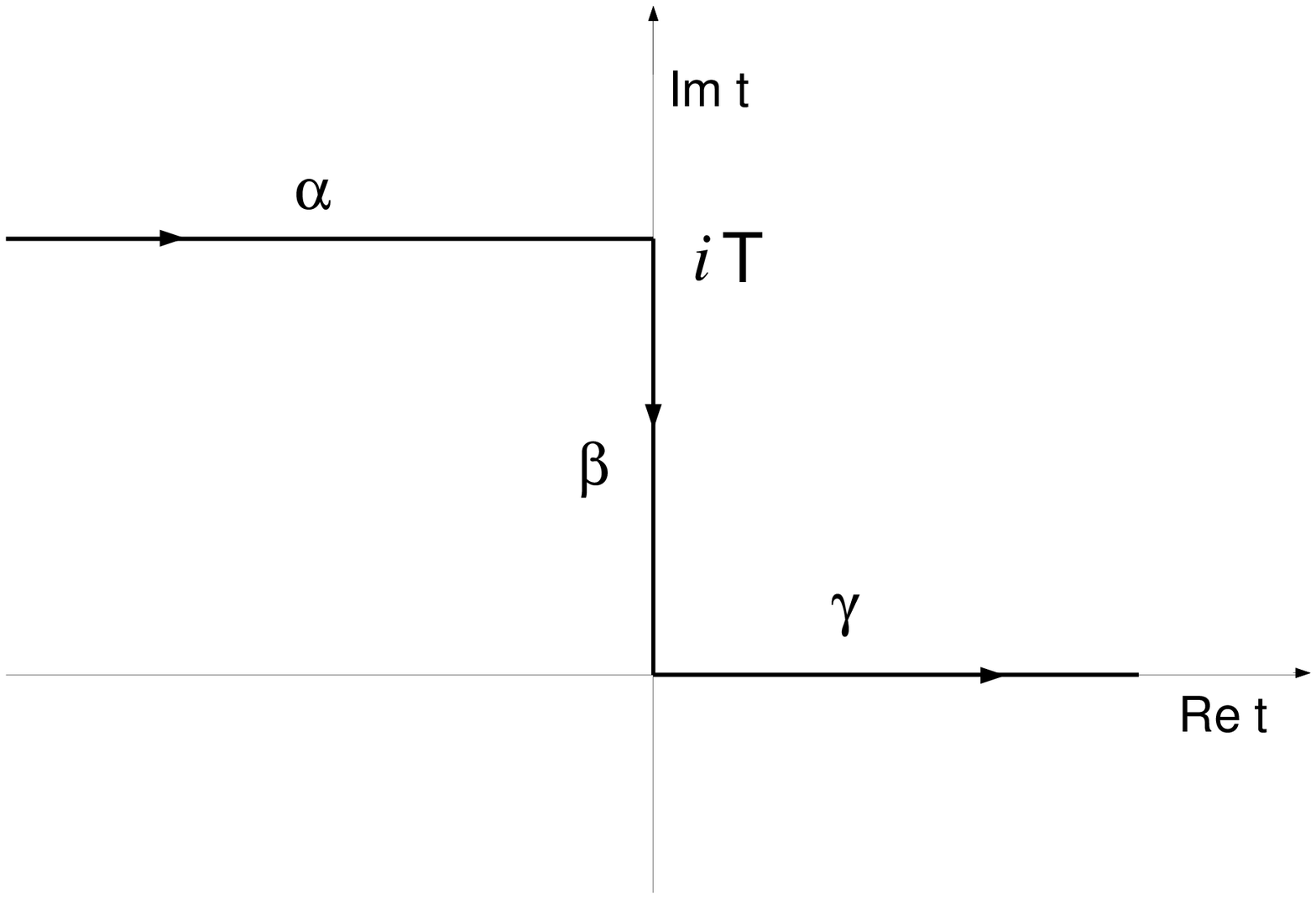}

Figure 1.~Complex time contour.
\vspace{4mm}

The boundary conditions are of special importance.  The solution
must be ``real'' along $\gamma$, where the asymptotic fields
for $t \to +\infty$ represent the final state (if the system
contains complex fields $\phi$ and $\bar\phi$, by reality we
mean that $\bar\phi = \phi^*$).
The condition on the particle number in the initial
state translates into the requirement that for very early times
in the evolution (i.e. asymptotically for $t \to -\infty + \imath
T$ along $\alpha$) the fields must reduce to a superposition of
normal modes of oscillation with amplitudes satisfying the equation

\be
\bar A(-{\bf k}) = e^{\theta} A^*({\bf k}) \ , \label{eq2}
\ee

\noindent
where $\theta$ plays the role of a chemical potential conjugate
to the particle number in the initial state.  It is clear from
Eq.~\ref{eq2} that for $\theta \ne 0$ the fields cannot be real
along $\alpha$.

Computationally, it is convenient to think of a time reversed
evolution by which an initial configuration at $t=0$ undergoes
a Euclidean evolution to $t=\imath T$ along the imaginary time
axis  (i.e. following the oriented segment $-\beta$), and then
a Minkowski evolution (along $-\alpha$) from $\imath T$ to
$-\infty + \imath T$. At $t=0$, for an $x$-axis discretized with
$N$ sites, we have $N$ independent real field coordinates and
$N$ pure-imaginary momentum components. The conditions on the
normal mode amplitudes given by Eq. \ref{eq2} amount to $N$
complex constraint equations, or $2N$ real equations.  Thus,
in principle, one could evolve the fields from an initial {\it Ansatz}
at $t=0$ and adjust the initial variables so that Eq.~\ref{eq2} is
satisfied. In practice, since the evolution equations along the imaginary
time axis are elliptic, one cannot perform a forward integration.
Rather, one must resort to some relaxation procedure or other global
algorithm by which one solves the evolution equations as a set of
simultaneous non-linear equations for all points of a space-time grid.
The situation is further complicated by the fact that with complexified
fields one cannot just minimize a Euclidean action integral.  We have
developed a ``second order'' formulation, by which we minimize a constraint
functional obtained from the modulus squared of the functions that must
vanish at all grid points (an earlier part of this study was done in
collaboration with Timothy Vaughan).  The formalism of lattice gauge
theory has been used to obtain a gauge invariant discretization. We
tested our procedure in the context of the $2D$ Abelian Higgs model
(one space and one time dimensions), where we found that it did reproduce
the expected Euclidean solutions, including solutions with multiple bounces
of the fields between two different topological sectors.

Another crucial ingredient of the calculation consists in extracting
the amplitudes of the normal modes of oscillations once the fields
reach the linear regime.  As the energy disperses in the course of
the Minkowski evolution, one expects that the system eventually reaches
a regime where the fields depart little from their vacuum values.
At this point the evolution becomes governed by the linearized field
equations and quantities like normal mode amplitudes and particle number
can be meaningfully defined.

Initially we studied the approach to the linear regime in the $2D$
Abelian Higgs system as a convenient simplified model which exhibits
the anomaly and the non-trivial topological properties required for
baryon number violation.  We found, however, that with this system
the decay of sphaleron like configurations gives origin to persistent,
localized excitations with an extremely slow rate of decay. These
excitations are reminiscent of the breather modes of $2D$ theories
with real scalar fields that exhibit soliton behavior. The Abelian
Higgs model has no soliton solutions in one spatial dimension, so
one expects the localized excitations we observed to eventually dissipate.
Still, their extremely long persistence made the computational extraction
of the normal modes amplitudes quite problematic.

This led us to consider the more realistic $4D$ $SU(2)$ Higgs system
where the energy should disperse much more readily because of the
higher dimensionality of space.  We implemented the evolution equations
for the spherically symmetric configurations studied by Ratra and Yaffe
\cite{ry88}. The spherically symmetric configurations have a residual
Abelian gauge symmetry and are described by a two component $U(1)$
gauge potential $a_{\mu}(r,t)$, and two complex Higgs-like fields
$\chi(r,t)$ and $\phi(r,t)$ of charge $1$ and $1/2$ respectively
($\chi$ parameterizes
the transverse components of the original $4D$ gauge potential and
$\phi$ the original Higgs field). The decay of the sphaleron in this
system has been studied by Hellmund and Kripfganz \cite{hk91} who
observed the onset of the linear regime.

\epsfxsize=75mm
\epsfysize=85mm
\epsfbox{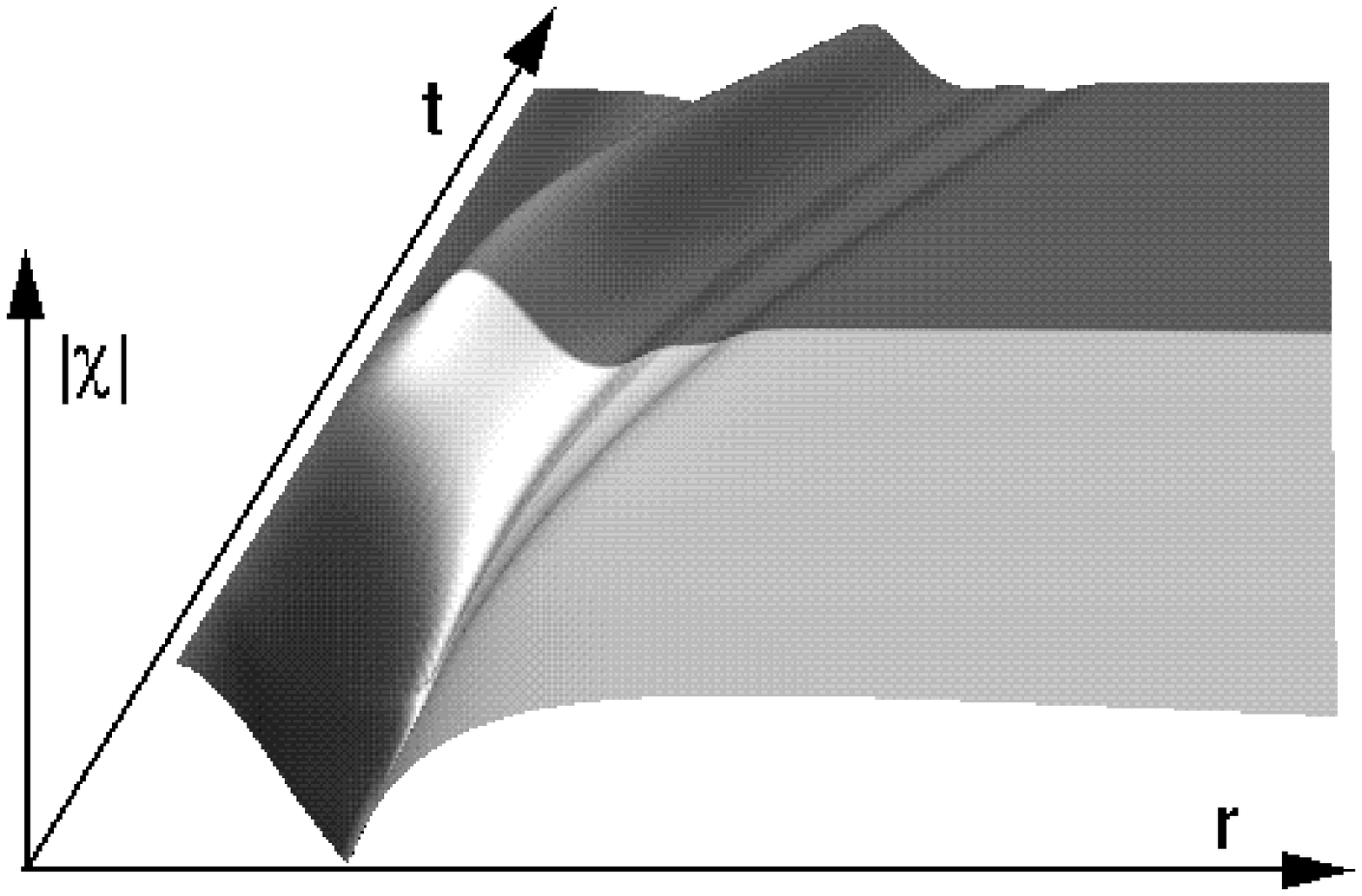}

\noindent
Figure 2.~Sphaleron decay in the four dimensional $SU(2)$ Higgs model:
evolution of the $\chi$ field. The values
of the phase of the complex field are coded by different shades of
grey.  When the evolution reaches the linear regime,
a gauge transformation,
indicated by the sudden change of shading, is performed to extract
the normal mode amplitudes.

\vspace{6mm}

We have developed code that simulates the Minkowski evolution of the
above system in the $a_0=0$ gauge.  Fig.~2 illustrates the decay
of a sphaleron after a small initial perturbation and the eventual
onset of the linear regime. We have calculated the normal modes
both numerically, as eigenmodes of the discretized system, and
by solving the continuum equations of motion analytically. There
are $4$ independent set of modes, corresponding to $3$ massive
gauge bosons and $1$ massive scalar boson.  In Fig.~3 we display
the time evolution of the total particle numbers of the $4$ sets of
modes.  The constancy of these observables at late time is a good
indication that the system has indeed reached the linear regime.
\vspace{2mm}

%
%
\epsfxsize=70mm
\epsfysize=85mm
\epsfbox{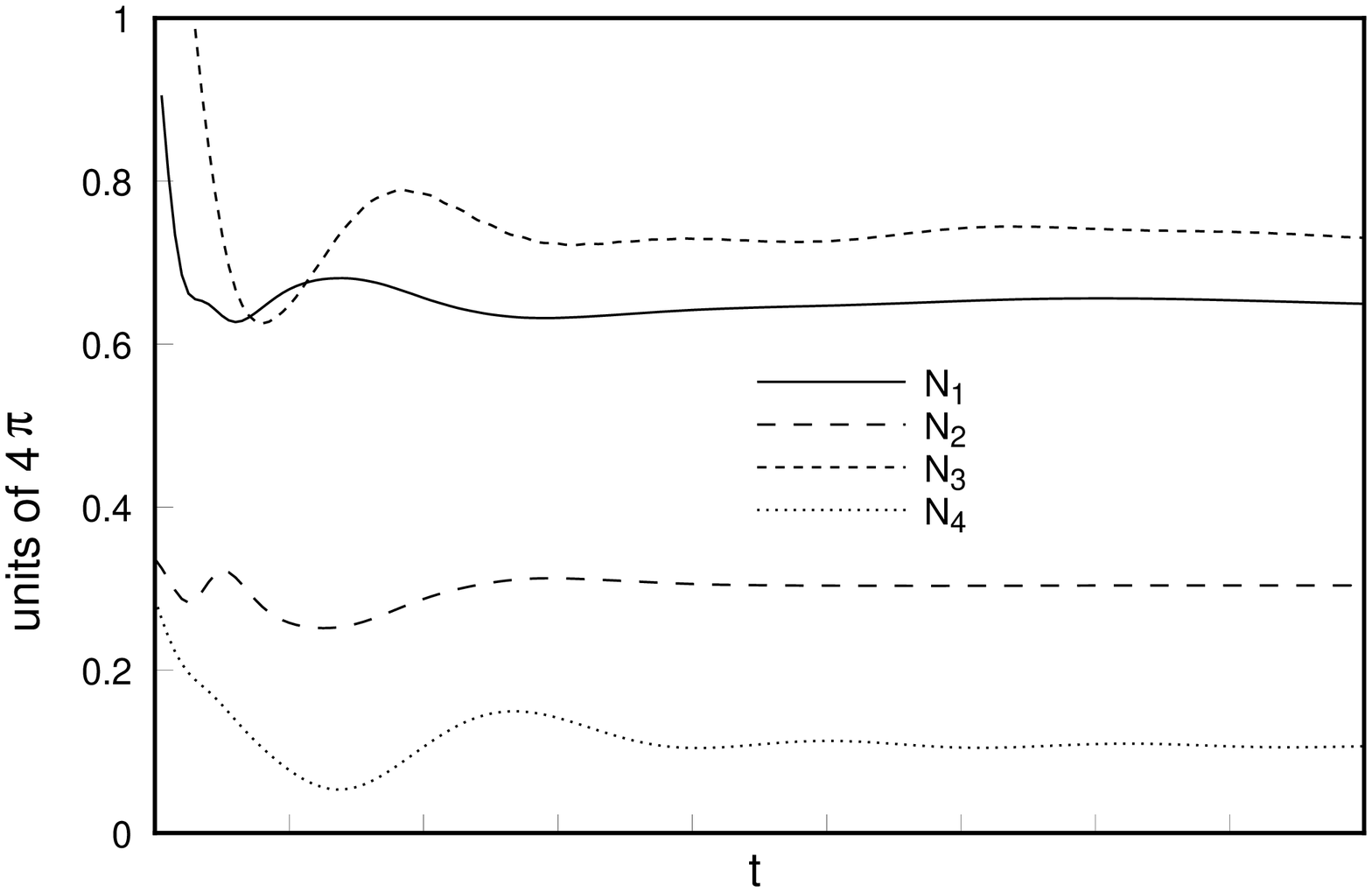}

\noindent
Figure 3.~Sphaleron decay in the four dimensional $SU(2)$ Higgs model:
behavior of the particle number in the four normal modes of oscillation
of the linearized system as function of time.

\vspace{10mm}

The identification of the normal-mode amplitudes also opens the possibility
of studying the induced violation of baryon number in the classically
allowed region with energy above the sphaleron barrier.  In this
region transitions that change the topology are not suppressed by a
barrier penetration factor, but the question remains open as to whether
they can occur in processes with an exclusive two-particle initial
state. Semiclassical methods can be applied again (cf.~Ref.~\cite{rs94})
and the analysis is somewhat simpler because the detour along a complex
time contour is no longer required.  Work on this problem is also in
progress.

\vspace{6mm}
\noindent {\Large \bf Acknowledgements}
\vskip2mm
We wish to thank Valery Rubakov for a conversation which gave
origin to this project, Andy Cohen and Peter Tinyakov for very useful
discussions, and Timothy Vaughan for participating in the early stages
of this work. This research was supported in part under DOE grant
DE-FG02-91ER40676 and NSF grant ASC-9405031.

\end{document}